\documentclass[aps,prl,twocolumn,superscriptaddress,showpacs]{revtex4}
\usepackage{graphics}
\usepackage{subfigure}
\usepackage{longtable}
\usepackage{times}
\usepackage{graphicx}
\usepackage{amsmath}
\usepackage{amsbsy}
\usepackage{dcolumn}
\usepackage{bm}

\def\LOFFA{{LaO$_{1-x}$F$_x$FeAs}}

\def\k{{\bf k}}

\begin{document}

\preprint{}

\title{Determining gap nodal structures in Fe-based superconductors:
angle-dependence of the low temperature specific heat in an applied magnetic field }

\author{S. Graser}
\affiliation{Department of Physics, University of Florida,
Gainesville, FL 32611, U.S.A.} \email[Corresponding author,
e-mail: ]{graser@phys.ufl.edu}
\author{G.R. Boyd }
\affiliation{Department of Physics, University of Florida,
Gainesville, FL 32611, U.S.A.}

\author{Chao Cao}
\affiliation{Department of Physics, University of Florida,
Gainesville, FL 32611, U.S.A.} \affiliation{Quantum Theory
Project, University of Florida, Gainesville, FL 32611, U.S.A.}

\author{Hai-Ping Cheng}

\affiliation{Department of Physics, University of Florida,
Gainesville, FL 32611, U.S.A.} \affiliation{Quantum Theory
Project, University of Florida, Gainesville, FL 32611, U.S.A.}

\author{P. J. Hirschfeld}
\affiliation{Department of Physics, University of Florida,
Gainesville, FL 32611, U.S.A.}

\author{D. J. Scalapino}

 \affiliation{Department of Physics,
University of California, Santa Barbara, CA 93106-9530 USA}

\date{\today}

\begin{abstract}
Since the discovery of  high-$T_c$ \LOFFA\, and other such systems
based on FeAs layers, several proposals have been made for the
superconducting order parameter  $\Delta_\k$, on both
phenomenological and microscopic grounds.  Here we discuss how the
symmetry of $\Delta_\k$ in the bulk can be determined, assuming that
single crystals will soon be available.  We suggest that a
measurement of the dependence of the low temperature specific 
heat on the angle of a magnetic
field in the FeAs plane is the simplest such method, and calculate
representative specific heat vs. field angle oscillations  for
the various candidate states, using a phenomenological band
structure fitted to the DFT Fermi surface.

\end{abstract}

\pacs{74.70.-b,74.25.Ha,74.25.Jb,74.25.Kc}

\maketitle

The recent discovery of superconductivity with onset temperature
of 26K in \LOFFA\ \cite{LOFA_JACS} was followed rapidly by the
development of materials with $T_c$ up to
$\sim$50K\cite{NLWang1,HHWen1,HHWen2,Jin,HHWen3,NLWang2,NLWang3,HHWen4,ChenXH},
which possess a similar structure but where  La has been replaced
by Sm, Pr, Nd or Ce. Common to all such materials is an
electronically layered structure, where according to electronic 
structure theories a rare-earth oxide layer where F substitutions 
for O dope an FeAs layer. The iron atoms are arranged in a simple 
square lattice, separated by arsenic atoms above and below this 
plane. The FeAs complex provides the bands at the Fermi level,
and the Fermi surface consists of sheets around
the $\Gamma$ point and the $M$ point of the Brillouin
zone\cite{LDA0,Xuetal,Mazin:2008,Cao:2008,LDA4,Haule:2008}.

Beyond this general initial consensus on the commonalities of the
different materials, electronic structure calculations differ on
the details of the ground state and the band structure near the
Fermi level.  Both paramagnetic and antiferromagnetic (sublattice
and linear SDW) ground states have been reported, with some
authors claiming that the system is close to a Mott transition and
also possibly to a ferromagnetic state.  Crude support for the
proximity of  competing magnetic states is provided by the known
helimagnetism in the layered iron monoarsenide system.

Within weeks of the discovery of the \LOFFA~systems, theoretical
analyses of various possibilities for the mechanism of
superconductivity and the symmetry of the superconducting order
parameter have
appeared\cite{Mazin:2008,Xuetal,Kuroki:2008,XDai,Hanetal}.
Eliashberg style calculations based on density functional theory
(DFT) determination of electron-phonon coupling
constants\cite{Phonon} suggest that conventional electron-phonon
interactions are not sufficient to generate the observed transition 
temperatures.  Thus several authors have discussed electronic pairing
mechanisms of the spin fluctuation
type\cite{Mazin:2008,Xuetal,Kuroki:2008,XDai,Hanetal}, but
disagree about the symmetry of the ground state, apparently
because of the details of the electronic structure used as an
input to the calculation. Given the past history of theoretical
approaches to unconventional superconductors, it may be some time
before a consensus on the correct microscopic approach is forged.

 In the intervening
period, it would clearly be useful to have some information on the
symmetry of the order parameter to guide such theoretical
discussions.  Evidence for nodes in the order parameter has
already been provided by point contact tunnelling\cite{Shan},
which has reported a zero bias state in a series of relatively
high-transparency junctions, and specific heat measurements in a
magnetic field $H$\cite{Muetal}, which indicate a $C_V/T\sim
\sqrt{H}$ term similar to that predicted by Volovik for a $d$-wave
(or, more generally, nodal) superconductor.  Because the current
experiments have been performed on powdered samples, however, the
distribution on the Fermi surface of order parameter nodes, which
could provide some information on the symmetry of the pair state
is not yet determined.  In addition, the point contact
measurements probe only the superconducting state at the surface,
whereas ideally one would prefer to extract information on the
bulk superconducting state. When single crystals are produced, it
will be possible to perform what is possibly the simplest bulk
probe of the distribution of gap nodes, a measurement of the
specific heat of a sample in the presence of a field in the FeAs
plane as a function of its angle relative to the crystal axes. The
superflow field in the vortex state of the type-II superconductor
is known to ``Doppler shift" the energies of quasiparticles,
changing their local occupation and giving rise to a residual
density of states\cite{volovik} which depends on the angle the
field makes with the nodes\cite{volovik2,VHCN}.   This method of
``nodal mapping" was proposed in the context of the cuprates,
where the experiment is difficult due to the large phonon
background, but has found more fruitful application in lower-$T_c$
materials\cite{Matsudareview}.

In this paper we calculate the specific heat oscillations with
magnetic field angle to be expected in the presence of a variety
of candidate superconducting pair states.  Rather than tie
ourselves to any particular microscopic electronic structure
calculation, we use a phenomenological two-band model \cite{Raghu}
which captures the essential qualitative features of the bands 
near the Fermi surface. We find that various
extended-$s$ like states can be  distinguished from, e.g. $d$-wave
or $p$-wave like states by the positions of their nodes.  There
are also cases, however, where nodes lie in positions on the Fermi
surface where $\k_n$ and the Fermi velocity ${\bf v}_F$ are not
parallel. In this case the minimum of the specific heat does not
correspond precisely to the nodal position and the
structure of the set of minima must be examined in detail.

{\it Effective band structure.} The crystal structure of LaOFeAs
consists of alternating layers of FeAs and LaO. Density functional
theory (DFT) calculations show that the energy bands crossing the
Fermi level can be assigned to the Fe $3d$ and the As
$4p$-orbitals \cite{LDA0,Xuetal,Mazin:2008,Haule:2008,Cao:2008,LDA4,Kuroki:2008}.
Thus, to describe superconducting properties we can consider the
LaO layers mainly as spacing layers and possible charge
reservoirs. The FeAs layers can be further subdivided into a 
square Fe lattice with an Fe-Fe spacing $d_{Fe-Fe}=2.82$ \AA~and
an As square lattice displaced by a vector $(1/2,1/2)$ in the
$x$-$y$-plane to a position in the center of the Fe squares.
Additionally, the As atoms are displaced alternately above or below
the Fe plane leading to a pyramidal Fe-As configuration. Due to the
alternating sign of the As $z$-displacements the primitive unit
cell of LaOFeAs contains two Fe and two As atoms. The axes of the
corresponding Brillouin zone (BZ) are aligned in the next 
nearest neighbor Fe-Fe direction and the BZ has a size of 
$2 \pi/a \times 2\pi/a$.
However, due to the high degeneracy of the two As positions we
will treat an effective model consisting of a smaller unit cell having
only one Fe and one As atom. This leads to a larger effective BZ 
that has axes that are aligned to the nearest neighbor Fe-Fe direction. In this
case the {\it real} BZ occupies a diamond shaped region within the
{\it effective} BZ.

\begin{figure}
\includegraphics[width=.8\columnwidth]{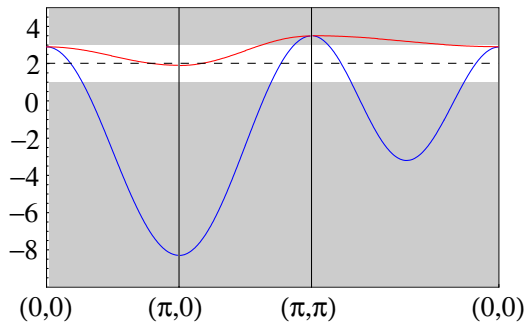}
\caption{(Color online) The two hybridized bands of our model. The
white region depicts the narrow energy window around the Fermi
energy (dashed line) where the model reproduces the
semiquantitative aspects of the LDA band structure.  The momenta
refer to points in the 2D effective large Brillouin zone (see Fig.
\ref{FS_fit}). } \label{BS}
\end{figure}

Since the thermodynamic properties of the superconducting state
are governed by low lying quasiparticle excitations, we  restrict
our considerations in the following to the region where the energy
bands cross the Fermi level, forming the different Fermi surface
sheets in the BZ. To simplify the rather complicated band
structure, we use a two band model that takes only the iron
$d_{xz}$ and $d_{yz}$ orbitals into account \cite{Raghu}.
Here the basic symmetry of the hopping parameters is determined
from the direct overlap of the Fe $d$ orbitals as well as from 
the hopping mediated by the As $p$ orbitals. The model 
neglects contributions from other orbitals, e.g. hybridization
due to the other Fe $d$ orbitals, and the hopping parameters
are adjusted to give the generic form of the Fermi surface
sheets determined by bandstructure calculations.

\begin{figure}
\includegraphics[width=.8\columnwidth]{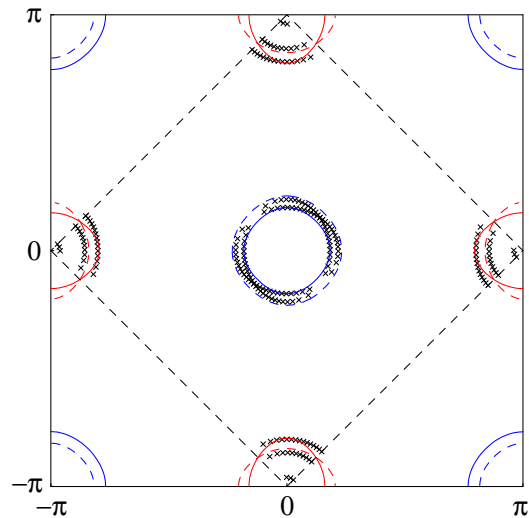}
\caption{(Color online) The different FS sheets in the large
effective BZ calculated within our two-band model. The two
hybridized bands result in two sets of Fermi surface sheets,
centered around the $\Gamma$ point (blue) and the $M$ point (red)
of the real BZ (dashed black line). Backfolding of the large into
the small BZ produces the dashed sheets in the small zone. The
black crosses show the FS of the paramagnetic ground state
determined by DFT \cite{Cao:2008}.} \label{FS_fit}
\end{figure}

Calculating the hopping from the direct overlap of the Fe $d_{xz}$
and $d_{yz}$ orbitals as well as the effective hopping in second
order perturbation theory on the path Fe-As-Fe, taking the As
$p_x$, $p_y$ and $p_z$ into account, leads to a tight binding
Hamiltonian with nearest and next-nearest neighbor hopping between
the same orbitals and a next-nearest neighbor exchange hopping
between the two bands. Due to the choice of the orbitals there are
different nearest neighbor hopping values $t_1$ and $t_2$ for
hopping in the $x$ and $y$ directions in one band which are
interchanged in the other band. The intraband next nearest
neighbor hopping $t_3$ is the same for both bands and both
directions, while the interband hopping $t_4$ has a different sign
for the $(1,1)$ compared to the $(1,-1)$ direction. After the
usual Fourier transformation we can write the intraband energies
in momentum space as
\begin{eqnarray}
\epsilon_{11}  &=& -2 t_1 \cos k_x -2 t_2 \cos k_y - 4 t_3 \cos k_x \cos k_y \\
\epsilon_{22}  &=& -2 t_2 \cos k_x -2 t_1 \cos k_y - 4 t_3 \cos
k_x \cos k_y
\end{eqnarray}
and the interband exchange energy is
\begin{equation}
\epsilon_{12} = \epsilon_{21}  =  - 4 t_4 \sin k_x \sin k_y
\end{equation}
Taking the hybridization of the two bands into account 
one finds for the two bands
\begin{eqnarray}
\epsilon^{\alpha}  &=& \frac{1}{2} \left(
\epsilon_{11}+\epsilon_{22} -
\sqrt{(\epsilon_{11}-\epsilon_{22})^2
                                     +4 \epsilon_{12}^2} \right)\\
\epsilon^{\beta}  &=& \frac{1}{2} \left(
\epsilon_{11}+\epsilon_{22} +
\sqrt{(\epsilon_{11}-\epsilon_{22})^2
                                     +4 \epsilon_{12}^2} \right)
\end{eqnarray}
The hopping parameters $t_i$ and the chemical potential $E_F$ 
can be used to fit the FS of the paramagnetic ground state found within the DFT
calculations \cite{Cao:2008}. We find a reasonable agreement of
the FS for the following values $t_1 = -1.2$, $t_2 = 1.35$, $t_3 =
-0.8$, $t_4 = -0.8$ and $E_F=2$.  These values lead to the
bandstructure shown in Fig. \ref{BS}.

\begin{figure}
\includegraphics[width=.8\columnwidth]{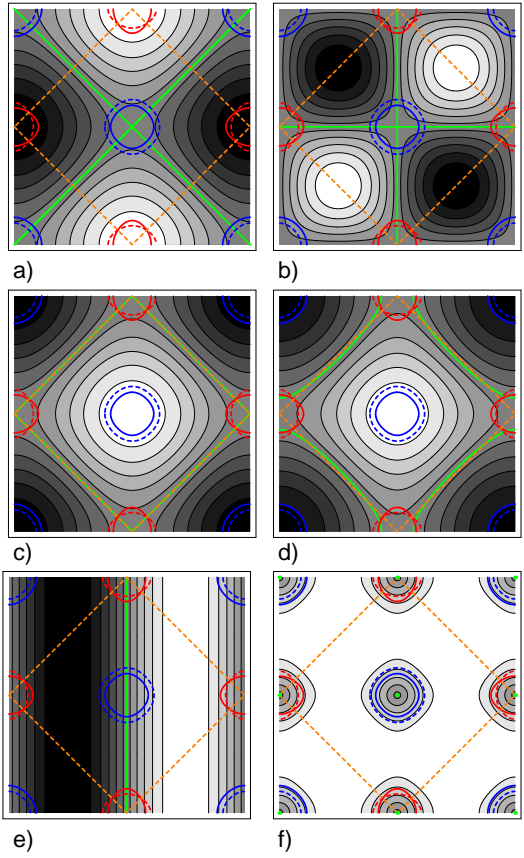}
\caption{(Color online)  Candidate order parameter states
considered in this work.  a) $d_{x^2-y^2}$ state; b) $d_{xy}$
state; c) extended-$s$ state from Ref. \cite{Mazin:2008} ; d)
generalized extended-$s$; e) $p_x$ state ($\Delta_\k\propto \sin
k_x$); f) $p_x+ip_y$ state ($\Delta_\k=\sin k_x +i \sin k_y$)~\cite{Xuetal}. 
The dashed orange line denotes the small Brillouin zone and the green
line denotes locus of gap nodes. } \label{Gaps}
\end{figure}

{\it Order parameters.}  Because of the small coherence length of
${\cal O}$(30 \AA), these systems are strongly type-II and it may
be appropriate to think of the range of the pairing interaction as
being very short, of order the lattice spacing.  Generally
speaking, order parameters involving pairing on nearest neighbor
sites are also those which have been proposed for these systems.  We
therefore consider as representative candidates the states listed
in Fig. \ref{Gaps}, proposed by various authors, beginning with
nearest-neighbor a) $d_{x^2-y^2}$ state ($\Delta_\k\propto\cos k_x
-\cos k_y$)\cite{Hanetal} ; b) $d_{xy}$ state $(\Delta_\k \propto
\sin k_x \sin k_y)$; and   c) extended $s$-wave state ($\Delta_\k
\propto \cos k_x +\cos k_y$) \cite{Mazin:2008}, and e) a $p_x$-wave
state.   The extended $s$-wave state shown in c) changes sign on
the Fermi surface of the model system, as seen in the Figure.  On
the other hand, its nodes are located at the 45$^\circ$ directions
relative to the sheet center at the M point, which is not generic
for a state with $s$ ($A_{1g}$) symmetry. We therefore show in
Fig. \ref{Gaps}d) the result of adding to this state a higher
order $s$-harmonic $\Delta_\k \propto (1-\gamma)(\cos k_x +\cos k_y) +
\gamma ( \cos 2k_x +\cos 2k_y)$ with $\gamma=0.05$.
For example, the RPA spin
fluctuation calculations of Kuroki et al. appear to lead to a more
general extended $s$-wave state. Similarly, it can be seen in Fig.
\ref{Gaps}d) that the points where the nodes cross the Fermi
surface sheets are away from the 45$^\circ$ directions (relative
to the center of the sheet on the zone face). Finally, we show in
Fig. \ref{Gaps}f) the nodeless $p_x+ip_y$ state proposed by Xu et
al. \cite{Xuetal}.

{\it Specific heat.} To get a qualitative understanding of the
specific heat oscillations as a function of the rotation angle of
an in-plane magnetic field at low temperature it is sufficient to
study the spectrum of low energy excitations. To calculate the
spectrum in the vortex state we want to follow a semiclassical
approach, that neglects the core states and considers only the
shift of the quasiparticle energies of the extended nodal states
in the presence of a magnetic field. Following \cite{VHCN} 
we approximate the vortex lattice using a circular unit cell 
with radius $R$ and winding angle $\beta$. Then the Doppler
shifted quasiparticle energy is
\begin{equation}
\delta E^{(i)} = m {\bf v}_F^{(i)} {\bf v}_s = \frac{E_H}{\rho}
                \left(\hat{v}_{F,y}^{(i)} \cos \alpha - \hat{v}_{F,x}^{(i)} 
                \sin \alpha \right) \sin \beta
\end{equation}
Here ${\bf v}_F^{(i)}$ denotes the Fermi velocity on band $i$,
${\bf v}_s$ is the gauge invariant expression of the quasiparticle
flow field around the vortex core and $\alpha$ is the angle
between the magnetic field and the $x$-axis of our coordinate
system. The dimensionless radial variable $\rho = r/R$ and 
$\hat{v}_{F,x/y}^{(i)}$ are the components of the Fermi
velocity calculated from $\nabla \epsilon_k^{(i)}$, normalized by
a Fermi surface averaged value of $v_F^{(i)}$. The energy scale
$E_H$ associated with the Doppler shift is
\begin{equation}
E_H^{(i)} = \frac{a}{2} \tilde{v}_F^{(i)} \sqrt{\pi H/\Phi_0}
\end{equation}
where $a$ is a geometric constant characteristic of the vortex
lattice, $\Phi_0$ is the flux quantum, and $\tilde{v}_F^{(i)}$ is 
an averaged Fermi velocity on
band $i$ determined from DFT calculations. This procedure of
normalizing the Fermi velocities prevents us from overestimating
the differences in the energy gradients at the Fermi level of the 
simplified two-band model. Using the
Doppler shifted energy in a BCS-like density of states we can
calculate the low energy spectrum as
\begin{equation}
N_0^{(i)} = \mathrm{Re} \left\langle \left\langle \frac{
|\delta E^{(i)}|}
 {\sqrt{\left(\delta E^{(i)} \right)^2 - \left(\Delta_k /E_H^{(i)} \right)^2 \rho^2}}
\right\rangle_{H} \right\rangle_{FS}
\end{equation}
where the angular brackets denote an average over the vortex cell
($H$) and over the Fermi surface ($FS$), respectively. The integral
over the vortex cell can be done analytically
leading to
\begin{equation}
N_0^{(i)} (\alpha)= \left\langle \mathrm{min} \left[ 1,
(E_H^{(i)}/\Delta_k)^2 \left(\hat{v}_{F,y}^{(i)} \cos \alpha -
\hat{v}_{F,x}^{(i)} \sin \alpha \right)^2 \right]
\right\rangle_{FS}
\end{equation}
The last average is to be performed over the different Fermi
surface sheets in the unfolded Brillouin zone leading to
an oscillation of the low energy spectrum as a function of the
angle of the applied magnetic field. These oscillations can be
directly determined by low temperature thermodynamic measurements,
like specific heat or the thermal conductivity.

\begin{figure}
\begin{center}
\includegraphics[width=.8\columnwidth]{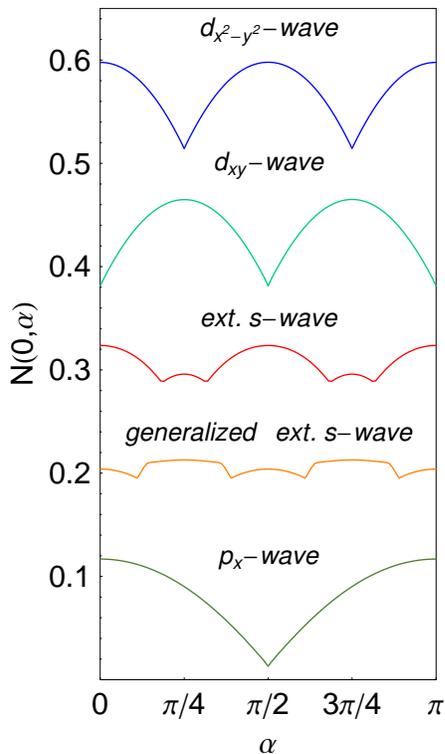}
\end{center}
\vspace{-0.5cm}
\caption{(Color online) Residual density of states
$N(\omega=0,{\bf H})$ vs. $\alpha$, the angle $\bf H$ makes with
the $x$-axis for the different superconducting states shown in Fig.
\ref{Gaps}. $N(\omega=0,{\bf H})$  is proportional to the 
linear specific heat coefficient at low temperature.
Note all curves have been offset by a constant amount
for clarity.} \label{N0alpha}
\end{figure}

In Fig. \ref{N0alpha}, we show the residual angle-dependent
density of states, or linear specific heat coefficient as a
function of field angle $\theta$.  For the most part, one expects
fairly straightforward generalizations of the results for a
circular Fermi surface\cite{VHCN}, as seen for the $p_x$ and
$d$-symmetry states: minima in the specific heat at low
temperatures $T\ll E_H^{(i)}$ lie at the expected
nodal positions.  The nodeless $p_x+ip_y$ state produces no
Volovik effect, is therefore not plotted in Fig. \ref{N0alpha}
and is apparently not a candidate for the Fe-based materials.  In the
extended-$s$ cases, some interesting points arise.  It is seen
from Fig. \ref{Gaps}c) that in the simple extended-$s$ case, the
nodes are located along the 45$^\circ$ directions.  Nevertheless
the minima in Fig. \ref{N0alpha} are slightly displaced
symmetrically with respect to these nodes; this is due to the fact
that the M sheets are elliptical, with the consequence that the
Fermi velocities are not parallel to nodal $\k_n$ measured from
the sheet center.  When there are higher harmonics, such as in the 
generalized extended $s$-wave case, the nodes themselves 
actually are displaced from the 45$^\circ$ directions.  
Thus a measurement of this kind
can identify an extended $s$ state by the displacements of the
minima, but a direct correspondence with the nodal positions
requires a precise knowledge of the underlying Fermi surface.

{\it Conclusions.} In this paper, we have proposed that the
measurement of specific heat oscillations as a function 
of the magnetic field
angle in the FeAs plane of the new iron-based superconductors
could be the simplest and most straightforward bulk determination
of gap symmetry, once single crystals or highly oriented powders
are available. To simplify the calculation, we used an
effective two-band model with parameters chosen to reproduce the
DFT Fermi surface.  We then calculated the low-temperature
linear term in the specific heat to be expected as a
function of field angle for a variety of candidate states.  The
elliptical Fermi surface pockets near the M points introduce some
interesting complications in the problem relative to the usual
picture of $C_V({\bf H})$ oscillations over the field angle.


\begin{acknowledgments}
This work is supported by DOE DE-FG02-02ER45995,
NSF/DMR/ITR-0218957 (HPC and CC), and DOE DE-FG02-05ER46236 (PJH). 
SG gratefully acknowledges support by the 
Deutsche Forschungsgemeinschaft. DJS acknowledges support
from the Center for Nanophase Material Science, ORNL.
\end{acknowledgments}

\end{document}